\documentclass[10pt,twocolumn]{article}
\usepackage{shock}
\usepackage{graphicx}
\usepackage{latexsym}
\usepackage{layout}
\usepackage{amssymb}
\usepackage{amsmath}

\newcounter{th}
\newenvironment{theorem}
{\refstepcounter{th} \textsc{Theorem \arabic{th}.}\it}

\newcounter{lm}
\newenvironment{lemma}
{\refstepcounter{lm} \textsc{Lemma \arabic{lm}.}\it}

\newcounter{co}
\newenvironment{corollary}
{\refstepcounter{co} \textsc{Corollary%\arabic{co}
.}\it}

\newcounter{df}
\newenvironment{defin}
{\refstepcounter{df} \textsc{Definition.}}

\newcounter{nt}
\newenvironment{remark}
{\refstepcounter{nt} \textsc{Remark.}}

\newenvironment{proof}
{\textsc{Proof.}}{$\Box$}

\newcommand{\R}{\mathbb{R}}
\newcommand{\cnst}{\mathrm{const}}
\newcommand{\nbl}{\nabla}
\newcommand{\lpl}{\Delta}
\newcommand{\id}{\mathrm{id}}

\newcommand{\pot}{\varphi}
\newcommand{\potv}{\psi}
\newcommand{\visc}{\nu}
\newcommand{\vel}{\mathbf{v}}

\newcommand{\ab}{\mathbf{a}}
\newcommand{\Ab}{\Xb_0}

\newcommand{\xb}{\mathbf{x}}
\newcommand{\tr}{x}
\newcommand{\Xb}{X}
\newcommand{\Tr}{\mathcal{X}}
\newcommand{\Trn}{\mathfrak{X}}

\newcommand{\Yb}{Y}
\newcommand{\try}{y}

\newcommand{\Trp}{\mathcal{P}}

\newcommand{\T}{\mathcal{T}}

\newcommand{\vb}{\mathbf{q}}
\newcommand{\ub}{\mathbf{u}}
\newcommand{\pb}{\mathbf{p}}
\newcommand{\Pb}{P}
\newcommand{\Vb}{Q}

\newcommand{\g}{\mathbf{g}}
\newcommand{\gn}{\mathfrak{g}}

\newcommand{\K}{\mathcal{D}}

\title{Matter evolution in Burgulence}

\author{
{\sc Ilya A. Bogaevsky}\\ \\
{\small Independent University of Moscow}\\
{\small Bolsho{\u\i} Vlas$'$evski{\u\i} per.\,11, Moscow 119002, Russia}\\
{\small \it E-mail: bogaevsk@mccme.ru}}

\date{}

\begin{document}

\maketitle

\begin{abstract}\noindent
In inviscid solutions of the forced Burgers equation the matter
accumulates in the shock discontinuities. We describe the limit
motion of particles everywhere including the shocks as the
trajectories of a discontinuous velocity field being a
generalization of the gradient of the limit potential. The latter
is not differentiable but satisfies some convexity properties
which guarantee the existence of the gradient. It turns out that
for such discontinuous gradient ordinary differential equations
there are natural existence, uniqueness, and continuity theorems.
These general results are applied for investigation of formation
and motion in plane of massive points which are interpreted as
various clusters in the adhesion model of the Universe.

\end{abstract}

\renewcommand{\thefootnote}{\fnsymbol{footnote}}
\footnotetext{Supported by RFBR-02-01-00655 and NSh-1972.2003.1.}

\section{Introduction}

The subject of this paper is the matter evolution in limit
potential solutions of the Burgers equation with vanishing
viscosity and external potential force. The Burgers equation is
just the Navier--Stokes equation without the pressure term -- its
theory is well described in the survey \cite{FB00}.

It is well known that in such inviscid potential solutions there
can be shocks, i.e. velocity discontinuities. They appear even if
the initial condition and the external force are {\it smooth} --
here and further this term means infinite differentiability.
Generically in this smooth case shocks are smooth hypersurfaces
with prescribed singularities. ``Generically'' means that other
singularities can be killed by arbitrarily small perturbation of
the smooth initial condition.

\begin{figure}[ht]
\begin{center}
\includegraphics{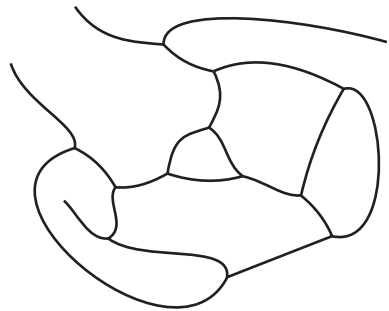}
\caption{Example of shock in plane} \label{ex}
\end{center}
\end{figure}

\begin{figure*}[t]
\begin{center}
\includegraphics{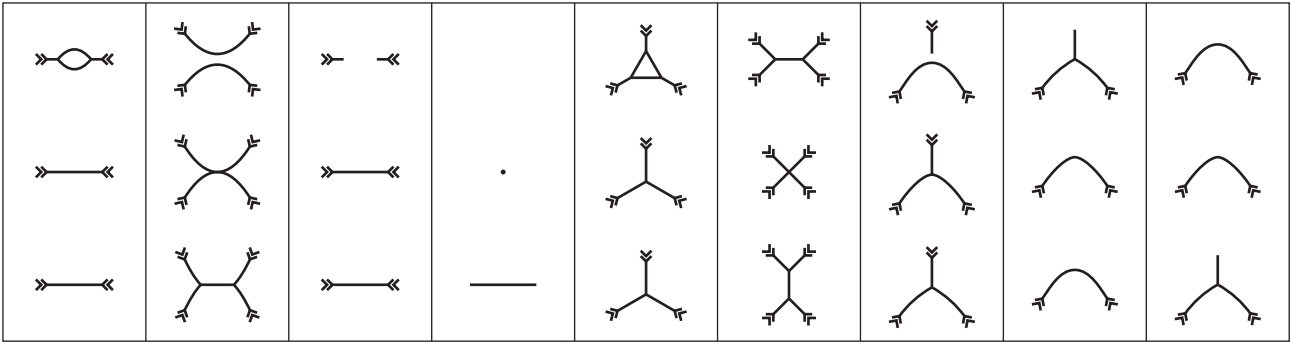}
\caption{Generic transitions of shocks in plane} \label{dd}
\end{center}
\end{figure*}

In plane such a generic shock is a smooth curve with triple nodes
and end points looking like shown in Figure\,\ref{ex}. It can
experience the transitions which are shown in Figure\,\ref{dd}.

Limit potential solutions of the Burgers equation describe motion
when the particles cannot pass through each other and adhere on
shocks. It happens because a particle cannot leave the shock. More
precisely, a particle trajectory ending outside of the shock lies
outside of the shock as well, but a trajectory beginning outside
of the shock can end on the shock. In other words, the matter
accumulates in the shocks where the density is infinite. This is
the so-called adhesion model of matter evolution in the Universe
describing the formation of cellular structure of the matter (see,
for example, \cite{GMS91}) -- the adhesion of particles is a
result of interaction between them described by the vanishing
viscosity.

But what is the motion of particles on shocks? This question is
answered by the present paper. Its main results are the following.

1) When the viscosity is positive the trajectory of any particle
is well defined because the velocity field is smooth. It turns out
there exists a limit of the trajectory as the viscosity vanishes.
Such limit trajectories describe the motion of particles in the
inviscid solution. The following uniqueness theorem is true: there
is only one limit trajectory beginning at a given point but there
can be a few limit trajectories ending at it. Of course, a few
trajectories can end only at a point of the shock.

2) How to find the limit trajectory of a given particle? It turns
out that the limit trajectory is a solution of the Cauchy problem
for a velocity field defined by the limit potential. The velocity
field is discontinuous but, nevertheless, the Cauchy problem has a
unique solution.

Besides, the above general results are applied for investigation
of formation and motion in plane of massive points (points with
positive mass) which are interpreted as various {\it clusters} in
the adhesion model of the Universe. It is natural to assume that
on the shock outside of its singularities the matter is
distributed with positive linear density and the nodes are massive
points. However, this idea is not correct!

\begin{figure}[ht]
\begin{center}
\includegraphics{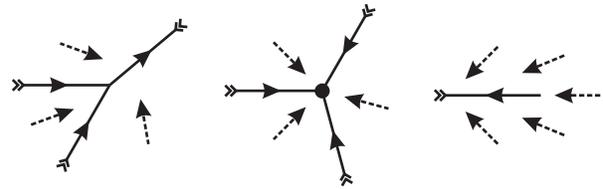}
\caption{Velocities of particles around acute node, growing
cluster at obtuse node, and end point} \label{ns}
\end{center}
\end{figure}

Indeed, the velocity field of the limit trajectories is smooth on
the shock outside of its singularities and a cluster cannot appear
there. But a cluster cannot appear at a node with an acute angle
too because the field of the relative (with respect to the node)
velocities looks like shown in Figure\,\ref{ns} on the left. So
particles pass through such a node and the matter does not
accumulate here. Otherwise, if all angles of a node are obtuse
then the matter is trapped at the node and it is a growing cluster
as it is shown in Figure\,\ref{ns} in the middle. Its right side
shows that a cluster cannot appear at an end point of the shock as
well.

\begin{figure}[ht]
\begin{center}
\includegraphics{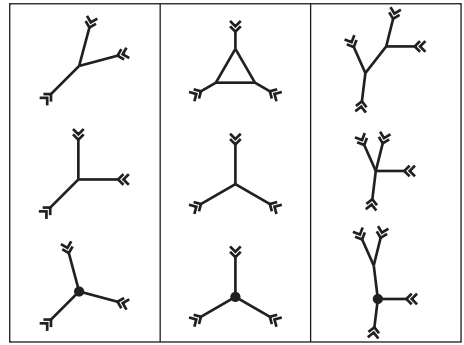}
\caption{Cluster is born and starts growing} \label{b}
\end{center}
\end{figure}

\begin{figure}[ht]
\begin{center}
\includegraphics{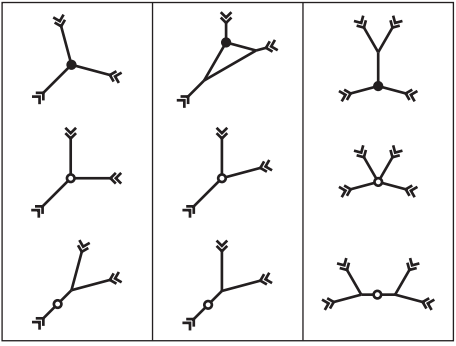}
\caption{Cluster stops growing and leaves node} \label{s}
\end{center}
\end{figure}

\begin{figure}[ht]
\begin{center}
\includegraphics{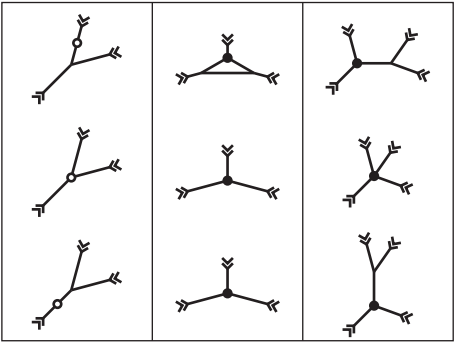}
\caption{Cluster travels through nodes} \label{p}
\end{center}
\end{figure}

\begin{figure}[ht]
\begin{center}
\includegraphics{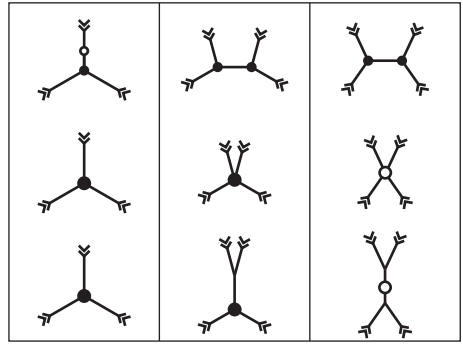}
\caption{Two clusters adhere} \label{i}
\end{center}
\end{figure}

So, a cluster is born when an acute node turns into an obtuse one
-- see the left transition in Figure\,\ref{b}. After the opposite
transformation the cluster stops growing and leaves the node --
this is the left transition in Figure\,\ref{s}. (In our figures
such stable clusters are shown by white disks, growing clusters --
by black disks.) A stable cluster travels along the shock and, in
particular, can pass through an acute node and be absorbed by a
growing cluster (the left transitions in Figures \ref{p} and
\ref{i} respectively).

The transitions in the middle and on the right from Figures
\ref{b}--\ref{i} show what happens with clusters when transitions
of shocks occur. Clusters can be involved only in the fifth and
sixth transitions from Figure\,\ref{dd} because all nodes of the
other ones are acute. Besides, generically a stable cluster cannot
come to a transition but can appear after it as shown in Figure
\ref{s} and \ref{i}.

I am very grateful to U.\,Frisch for calling my attention to the
problem as well as to him, J.\,Bec, M.\,Blank, K.\,Khanin,
R.\,Mohayaee, and A.\,Sobolevsky for fruitful discussions.

\section{Existence of limit trajectories}

So, we consider a material $d$-dimensional medium whose velocity
is potential and described by the Burgers equation with the
potential force term:
$$
%\begin{equation}
%\label{Burgers}
\left\{
\begin{array}{l}
\vel^\visc_t + (\vel^\visc \cdot \nbl{}) \, \vel^\visc = -\nbl{U}
+ \visc \lpl{\vel^\visc}\\
\vel^\visc = \nbl{\potv^\visc}\\
\potv^\visc(\xb,0) = \pot_0(\xb)
\end{array}
\right.
%\end{equation}
$$
where $\xb \in \R^d$ is a point of the medium, $\vel^\visc(\xb,t)$
is the velocity at the point $\xb$ at the time $t$, $\visc > 0$ is
the viscosity of the medium, $\nbl = (\partial_{x_1}, \dots,
\partial_{x_d})$ is the usual $\nbl$-operator in $\R^d$, and
$\lpl=\nbl \cdot \nbl$ is the Laplacian. The potential
$\potv^\visc$ of the velocity field $\vel^\visc$ is defined with
respect to a function of time which can be chosen so that the
following equation is satisfied:
\begin{equation}
\potv^\visc_t + \frac12 \nbl{\potv^\visc} \cdot \nbl{\potv^\visc}
+ U = \visc \lpl{\potv^\visc}. \label{brg}
\end{equation}

The force potential $U$ and the initial condition $\pot_0$ are
assumed to be {\it smooth}. (Everywhere in the present paper it
means infinite differentiability.) Let $\pot$ be the {\it limit
solution} as the viscosity vanishes.
\begin{equation}
\label{L} \pot(\xb,t)=\lim_{\visc \to 0} \potv^\visc(\xb,t).
\end{equation}
According to Theory of PDE, the potential $\potv^\visc$ is smooth
if $\visc > 0$ and $t \ge 0$. As it has been mentioned before, the
limit potential is continuous, but its gradient field can have
discontinuities (shocks).

We consider the periodical case. It means that all data -- the
given force potential $U$, the initial condition $\pot_0$, the
velocity field $\vel^\visc$, its potential $\potv^\visc$, and the
limit potential -- are assumed to be space-periodical. In other
words, we consider our equations on torus.

\medskip

\begin{remark}
The periodicity requirement is technical. Informally speaking, it
is needed to guarantee that nothing goes to the infinity and
nothing comes from the infinity for a finite period. Otherwise,
let this informal requirement be satisfied and we want to apply
our results in a finite domain of the space-time. Then we can
always consider our data as periodical with an enough large
period. Indeed, if we change them enough far we will not be able
to observe the influence in our finite domain.
\end{remark}

\medskip

Let $\try^\visc: [0, + \infty) \to \R^d$ be the trajectory
beginning at an initial point $\ab \in \R^d$, that is the solution
of the following Cauchy problem:
$$
\dot{\try}^\visc(t) = \nbl{\potv^\visc}(\try^\visc(t),t), \quad
\try^\visc(0) = \ab
$$
which has a unique solution because the right side of the ordinary
differential equation is a smooth space-periodical vector field.

\medskip

\begin{theorem}
\label{lmt} {\bf Existence:} For any initial point $\ab$ there
exists a limit trajectory $\tr: [0, + \infty) \to \R^d$
$$
\tr(t) = \lim_{\visc \to 0} \try^\visc(t), \quad \tr(0) =
\try^\visc(0) = \ab.
$$
The convergence is uniform on any segment $[0, T]$ and the limit
trajectory is continuous.

\medskip

{\bf Uniqueness:} If two limit trajectories pass through the same
point at some time then they coincide {\bf after} that time:
$$
\tr_1 (t_\ast) = \tr_2 (t_\ast) \quad \Rightarrow \quad \tr_1 (t)
= \tr_2 (t) \quad \forall \; t \ge t_\ast.
$$
(But they may not coincide {\bf before} the time: $\tr_1 (t) \ne
\tr_2 (t)$ for $t < t_\ast$.)

\medskip

{\bf Continuity:} The point $\tr(t)$ is a continuous function of
the time $t$ and the initial point $\ab$.
\end{theorem}

\medskip

\begin{corollary}
For any point $\xb_\ast$ and time $t_\ast \ge 0$ there is a limit
trajectory $\tr: [0, + \infty) \to \R^d$ passing through the point
at the time: $\tr(t_\ast) = \xb_\ast$.
\end{corollary}

\medskip

\begin{proof}
According to Theory of ODE, for any $\visc$ there exists a
trajectory $\try^\visc$ such that $\try^\visc(t_\ast) = \xb_\ast$
(because $\vel^\visc$ is smooth). Taking into account that our
torus is compact we can choose a sequence $\visc_n \to 0$ as $n
\to \infty$ such that the sequence $\try^{\visc_n}(0)$ converges
to a point $\ab_\ast$ as $n \to \infty$. For the limit trajectory
$\tr$ with the initial point $\tr(0) = \ab_\ast$ we get
$\tr(t_\ast) = \xb_\ast$.
\end{proof}

\section{Differential equation for limit trajectories}

\begin{theorem}
\label{vd} The derivative
$$
\pot^\prime_{\xb,t}(\vb,0) = \lim_{\lambda \to +0}{\frac {\pot
(\xb + \lambda \vb, t) - \pot(\xb,t)} {\lambda}}
$$
of the limit potential along any space direction $\vb$ exists and
can be presented as the minimum of linear functions:
\begin{equation}
\label{fd} \pot^\prime_{\xb,t} (\vb,0) = \min_{\pb \in k_{\xb,t}}
{\pb \cdot \vb}
\end{equation}
where $k_{\xb,t}$ is a compact set of momenta which depends on the
point $(\xb,t)$ of the space-time.

Besides, any limit trajectory from Theorem\,\ref{lmt} satisfies
the differential equation
$$
\tr^+(t) = \ub(\tr(t),t)
$$
where $\ub(\xb,t)$ is the center of the minimal ball containing
the set $k_{\xb,t}$
%from the formula (\ref{fd})
and the left side is the one-way derivative
$$
\tr^+(t) = \lim_{\lambda \to +0} \frac{\tr(t + \lambda) - \tr(t)}
{\lambda}.
$$
\end{theorem}

\medskip

\begin{remark}
The set $k_{\xb,t}$ consists of the limit velocities at the time
$t$ at points which are outside of the shock and tend to the point
$\xb$.
\end{remark}

\medskip

Theorem\,\ref{vd} is proved in Section\,\ref{tvd} and can be
briefly explained in the following way. When $\visc \ne 0$ the
velocity $\vel^\visc(\xb,t) = \nbl{\potv^\visc} (\xb, t)$ is the
solution of the following minimum problem:
$$
|\vb|^2/2 - {\potv^\visc}^\prime_{\xb,t}(\vb,1) \to \min_\vb.
$$
It turns out that this minimum principle remains valid for the
limit potential. Namely, the limit velocity $\ub(\xb,t)$ is the
solution of the same minimum problem for the limit potential:
\begin{equation}
\label{min} |\vb|^2/2 - \pot^\prime_{\xb,t}(\vb,1) \to \min_\vb.
\end{equation}
Proving this principle in Section\,\ref{gde} we do not use that
the potentials $\potv^\visc$ are solutions of the Burgers equation
-- it is only important that their second derivatives are
uniformly bounded above.

\medskip

\begin{remark}
The principle (\ref{min}) is not variational because there is no
an integral functional to be minimized by this principle. In fact,
it just generalizes the notion of a gradient for a some class of
non-smooth functions of $\xb$ and $t$ (see Section\,\ref{gde} for
details). But such the generalized gradient depends on the
behavior of the function at closed times after $t$. It does not
happen if the function is smooth -- then we get the usual gradient
defined completely by the first space derivatives of the function.
\end{remark}

\medskip

Let us show how the principle (\ref{min}) implies
Theorem\,\ref{vd}. It is well known that in the case when the
initial condition $\pot_0$  is smooth there is the so-called
minimum representation
$$
\pot(\xb,t) = \min_{\xi}{ \left\{ F(\xi,\xb,t) \right\} }
$$
where $F$ is a family of smooth solutions of the Hamilton--Jacobi
equation:
$$
F_t(\xi,\xb,t) + |\nbl_\xb {F} (\xi,\xb,t)|^2 / 2 + U(\xb,t) = 0
$$
depending smoothly on a parameter $\xi$. (In other words, $F$ is a
smooth function of its variables $\xi$, $\xb$, and $t$.) It
immediately implies that the derivative of the limit potential
along a direction of the space-time
$$
\pot^\prime_{\xb,t}(\vb,\tau) = \lim_{\lambda \to +0}{\frac {\pot
(\xb + \lambda \vb, t + \lambda \tau) - \pot(\xb,t)} {\lambda}}
$$
can be presented as the minimum of some solutions of the
Hamilton--Jacobi equation freezed at the point $(\xb,t)$:
\begin{equation}
\label{fdn} \pot^\prime_{\xb,t}(\vb,\tau) = \min_{\pb \in
K_{\xb,t}} {\left\{ \pb \cdot \vb - \tau |\pb|^2 /2 - U(\xb,t) \,
\tau \right\}}
\end{equation}
where $K_{\xb,t}$ is a compact set of momenta which depends on the
point $(\xb,t)$ of the space-time. Substituting here $\tau = 0$
and comparing with the formula (\ref{fd}) we get that the sets
$k_{\xb,t}$ and $K_{\xb,t}$ has the same convex hull.

Applying the principle (\ref{min}) we get that the value
$$
|\vb|^2/2 - \pot^\prime_{\xb,t} (\vb,1) =
$$
$$
= |\vb|^2 / 2 - \min_{\pb \in K_{\xb,t}} {\left\{ \pb \cdot \vb -
|\pb|^2 /2  - U(\xb,t) \right\}} =
$$
$$
= \max_{\pb \in K_{\xb,t}} {\left\{ |\pb - \vb|^2 /2 \right\}} -
U(\xb,t)
$$
attains its minimum at the center of the minimal ball containing
the set $k_{\xb,t}$ because the latter has the same convex hull as
the set $K_{\xb,t}$.

\medskip

\begin{remark}
The equivalence of Theorem\,\ref{vd} and the principle (\ref{min})
for the limit solutions of the Burgers equation has been observed
independently on the author of the present paper by K.\,Khanin and
A.\,Sobolevsky.
\end{remark}

\section{Proof of Theorem\,\ref{lmt}}

\begin{lemma}
For any $T > 0$ there exists a constant $C(T)$ bounding for all
$\visc > 0$ and $0 \le t \le T$ the second derivative of the
potential in any direction of the space-time
$$
\potv^\visc_{\Vb \Vb} \le C(T),
$$
where $\Vb=(\vb, \tau)$, $\vb = (q_1, \dots, q_d)$, $|\vb|^2 +
\tau^2 =1$. \label{bnd}
\end{lemma}

\medskip

\begin{proof}
This is the standard maximum principle for the second derivative
$\potv^\visc_{\Vb \Vb}$ that satisfies the equation
$$
\potv^\visc_{\Vb \Vb t} + \nbl{\potv^\visc_\Vb}\cdot
\nbl{\potv^\visc_\Vb} + \nbl{\potv^\visc} \cdot
\nbl{\potv^\visc_{\Vb \Vb}} + U_{\Vb \Vb} = \visc
\lpl{\potv^\visc_{\Vb \Vb}}
$$
which is a consequence of (\ref{brg}) and implies the inequality
$$
\potv^\visc_{\Vb \Vb t} + \nbl{\potv^\visc} \cdot
\nbl{\potv^\visc_{\Vb \Vb}} + B \le \visc \lpl{\potv^\visc_{\Vb
\Vb}}
$$
where
$$
B(T) =  \min {U_{\Vb \Vb}(\xb,t)}, \; |\Vb|=1, \; \xb \in \R^d, \;
0 \le t \le T.
$$
Then the following inequality holds:
$$
\eta_t + \nbl{\potv^\visc} \cdot \nbl{\eta} \le \visc \lpl{\eta},
\quad \eta = \potv^\visc_{\Vb \Vb} + B t.
$$
So, when the function $\eta$ attains its maximal value for $t \in
[0,T]$ we get the inequality $\eta_t \le 0$ which shows that the
maximal value can be attained only if $t=0$. Therefore,
$$
\eta - B t \le \max{\{\eta|_{t=0}\}} - B t
$$
that means
$$
\potv^\visc_{\Vb \Vb} \le \max{ \{ \potv^\visc_{\Vb \Vb}|_{t=0} \}
} - B t
% = \max{\{{\pot_0}_{\Vb \Vb}\}} - B t
$$
but $\max{ \{ \potv^\visc_{\Vb \Vb}|_{t=0} \} }$ is defined by the
initial condition $\pot_0$ and the force potential $U$ because
$$
\potv^\visc_{\Vb \Vb} = \potv^\visc_{\vb \vb} + 2 \tau
\potv^\visc_{\vb t} + \tau^2 \potv^\visc_{t t}
$$
where according to (\ref{brg}):
$$
\potv^\visc_{\vb t} = \visc \lpl{\potv^\visc_\vb} -
\nbl{\potv^\visc} \cdot \nbl{\potv^\visc_\vb} - U_\vb,
$$
$$
\potv^\visc_{t t} = \visc \lpl{\potv^\visc_t} - \nbl{\potv^\visc}
\cdot \nbl{\potv^\visc_t} - U_t,
$$
$$
\potv^\visc_t = \visc \lpl{\potv^\visc} - \frac12
\nbl{\potv^\visc} \cdot \nbl{\potv^\visc} - U.
$$
If, for example,
$$
C(T) = \max{\{\pot_{\Vb \Vb}\}} + |B(T)| T,
$$
we get the statement.
\end{proof}

\medskip

\begin{lemma}
For any $T > 0$ the convergence (\ref{L}) is uniform on $\R^d
\times [0,T]$. \label{uc}
\end{lemma}

\medskip

\begin{proof}
According to Lemma\,\ref{bnd} the functions
$$
C(T) \,(|\xb|^2 + t^2)/2 - \potv^\visc(\xb,t)
$$
are convex on $\R^d \times [0,T]$. Hence, their convergence as
$\visc \to 0$ is uniform on any compact subset according to
Theorem\,\ref{gp} from Section\,\ref{CF}. But the functions
$\potv^\visc(\xb,t)$ are space-periodical and their convergence is
uniform on $\R^d \times [0,T]$.
\end{proof}

\medskip

We are proving Theorem\,\ref{lmt}. Let $\alpha, \varepsilon
> 0$. Lemma\,\ref{uc} implies that for any sufficiently small
$\visc$ and $\visc_\ast$
$$
| \potv(\xb) - \potv_\ast(\xb) | < \varepsilon
$$
where $\potv(\xb) = \potv^{\visc}(\xb,t)$, $\potv_\ast(\xb) =
\potv^{\visc_\ast}(\xb,t)$, $\xb \in \R^d$, and $t \in [0,T]$. We
want to get a uniform upper bound for the square of the distance
between the corresponding trajectories:
$$
R (t) = |\try (t) - \try_\ast (t)|^2
$$
where $\try(t) = \try^{\visc}(t)$, $\try_\ast (t) =
\try^{\visc_\ast}(t)$,
$$
\try(0) = \ab, \quad \try_\ast(0) = \ab_\ast, \quad |\ab -
\ab_\ast| < \alpha.
$$
According to Lemma\,\ref{bnd}:
$$
\potv_\ast(\try) - \potv_\ast(\try_\ast) \le
\nbl{\potv_\ast}(\try_\ast) \cdot (\try - \try_\ast) + C \, |\try
- \try_\ast|^2 / 2,
$$
$$
\potv(\try_\ast) - \potv(\try) \le \nbl{\potv}(\try) \cdot
(\try_\ast - \try) + C \, |\try_\ast - \try|^2 / 2.
$$
Adding the inequalities we get:
$$
- 2 \varepsilon < - (\nbl{\potv}(\try) -
\nbl{\potv_\ast}(\try_\ast)) \cdot (\try - \try_\ast) + C \, |\try
- \try_\ast|^2,
$$
or
\begin{equation}
(\nbl{\potv}(\try) - \nbl{\potv_\ast}(\try_\ast)) \cdot (\try -
\try_\ast) < 2 \varepsilon + C \, |\try - \try_\ast|^2,
\label{cvx}
\end{equation}
that gives
$$
\dot{R}(t) < 4 \varepsilon + 2 \, C \, R(t).
$$
Solving the differential inequality and taking into account that
$R(0) < \alpha$, we get:
$$
R(t) < 2 \varepsilon \frac{e^{2 C t} - 1}{C} + \alpha \, e^{2 C
t},
$$
or in the special case $C=0$:
$$
R(t) < 4 \varepsilon t + \alpha.
$$
The inequalities give the required uniform upper bound.

\section{Convex functions}

\label{CF}

Let $M$ be a convex subset of an $m$-dimensional affine space. (It
means that for any two points of $M$ the segment connecting them
belongs to $M$ as well.) A function $f: M \to \R$ is called {\it
convex} if it satisfies the inequality
$$
f(\alpha \Xb + \beta \Yb) \le \alpha f(\Xb) + \beta f(\Yb)
$$
for all $\alpha, \beta \ge 0$ such that $\alpha + \beta = 1$ and
any points $\Xb,\Yb \in M$.
% The function $f$ is called \textit{concave} if the function $-f$
% is convex.
A smooth function is convex
% (concave)
if and only if its second derivative along any direction is
non-negative.
% (non-positive). The sum of a concave function and a smooth one is
% called an {\it almost concave} function.

It is well known -- see, for example \cite{RF} -- that
% (almost)
convex functions have many good properties:

\medskip

\begin{theorem}
\label{conv1} Let $M \in \R^m$ be an open convex subset and  $f: M
\to \R$ be
% an almost concave function or just a concave function.
a convex function.

\medskip

{\bf Continuity:} $f$ is continuous on $M$.

\medskip

{\bf Differentiability:} $f$ has a finite derivative along any
direction $\Vb$ at any point $\Xb \in M$:
$$
f^\prime_{\Xb}(\Vb) = \lim_{\lambda \to +0} \frac {f(\Xb + \lambda
\Vb) - f(\Xb)} {\lambda}.
$$
Moreover,
$$
f(\Xb + \Vb) = f(\Xb) + f^\prime_{\Xb}(\Vb) + o(|\Vb|), \quad \Vb
\to 0
$$
where the derivative $f^\prime_\Xb(\Vb)$ is a convex homogeneous
function of $\Vb$:
$$
f^\prime_\Xb (\lambda \Vb) = \lambda f^\prime_\Xb(\Vb), \quad
\lambda \ge 0
$$
and can be presented as
$$
f^\prime_\Xb (\Vb) = \max_{\Pb \in \K_\Xb(f)} {\Pb \cdot \Vb}
$$
where the set $\K_\Xb(f)$ is convex and consists of the
sub-differentials of the function $f$ at the point $\Xb$.

\medskip

{\bf Sub-differential boundedness:} The sub-differentials
$\K_\Xb(f)$ is uniformly bounded if $\Xb$ belongs to a compact
subset of $M$.
\end{theorem}

\medskip

This theorem is proved in \cite{RF} (see theorems 10.1 and 23.1).

\medskip

\begin{theorem}
Let $M \in \R^m$ be an open convex subset and $f^\visc: M \to \R$
be a family of convex functions depending on a parameter $\visc$.

\medskip

{\bf Uniform convergence:} If the family converges
$$
f_\ast(\Xb) = \lim_{\visc \to 0}{f^\visc}(\Xb)
$$
then the limit function $f_\ast$ is convex on $M$ and the
convergence is uniform on any compact subset of $M$.

\medskip

{\bf Uniform derivative boundedness:} If the family $f^\visc$ is
uniformly bounded on $M$ then the family of the sub-differentials
$\K_\Xb(f^\visc)$ is uniformly bounded if $\Xb$ belongs to a
compact subset of $M$. \label{gp}
\end{theorem}

\section{Gradient differential equations}

\label{gde}

Let $M \subset \R^m$ be an open convex subset; a potential $\pot:
M \to \R$ be the difference of a semi-definite quadratic form $e$
and a convex function $f$:
$$
\pot(\Xb) = e(\Xb) - f(\Xb), \quad \Xb \in M;
$$
$\T = \R^m$ be the tangent space to $M$ or the space of
velocities; $\T^\ast = {\R^m}^\ast$ be the cotangent space to $M$
or the space of momenta; and a Hamiltonian $h: \T^\ast \to \R$ be
a smooth convex function of momenta.

The derivative of $\pot$ can be written in the following form:
$$
\pot_\Xb^\prime(\Vb) = \min_{\Pb \in \K_\Xb}{\Pb \cdot \Vb}, \quad
\Pb \in \T^\ast, \quad \Vb \in \T
$$
where $\K_\Xb = \K_\Xb(\pot) \subset \T^\ast$ is a compact convex
set of momenta -- see Theorem\,\ref{conv1} for details. The
momenta from the set $\K_\Xb$ are called {\it sub-differentials}
of the potential $\pot$ at the point $\Xb$.

Let the convex Hamiltonian $h$ attain its minimal value on
$\K_\Xb$ at a point $\Pb_\Xb \in \K_\Xb$ of the convex set of the
sub-differentials of the potential $\pot$.

\medskip

\begin{defin}
1) \textit{Hamiltonian form:} The velocity
% \left. \frac{\partial h}{\partial \Pb} \right| _{\Pb
% = \Pb_\Xb =
$\Vb_\Xb = h^\prime_{\Pb_\Xb} \in \T$ is denoted by
$\nbl_h{\pot}(\Xb)$ and called the {\it $h$-gradient} of the
potential $\pot$ at the point $\Xb$. Here $h^\prime_{\Pb_\Xb}$ is
the differential of the Hamiltonian $h$ at the point $\Pb_\Xb$.

\medskip

2) \textit{Lagrangian form:} The $h$-gradient $\Vb_\Xb$ is the
minimum point of the function
$$
l (\Vb) - \pot_\Xb^\prime(\Vb), \quad l(\Vb) = \max_{\Pb} {
\left\{ \Pb \cdot \Vb - h (\Pb) \right\} }
$$
where $l$ is the Lagrangian being the Legendre transformation of
the Hamiltonian $h$.
\end{defin}

\medskip

In order to show that the Hamiltonian and Lagrangian forms are
equivalent, let us note that
\begin{equation}
\label{bp}
% \pot^\prime_\Xb \left( \nbl_h{\pot}(\Xb) \right) =
% \Pb_\Xb \cdot \nbl_h{\pot}(\Xb) \; \mbox{ or } \;
\pot^\prime_\Xb \left( \Vb_\Xb \right) = \Pb_\Xb \cdot \Vb_\Xb
\end{equation}
where $\Vb_\Xb = \nbl_h {\pot} (\Xb)$. Indeed,
$$
\min_{\Pb \in \K_\Xb} {\Pb \cdot \Vb_\Xb} = \Pb_\Xb \cdot \Vb_\Xb
$$
because $\Pb_\Xb$ is a minimum point of the smooth function $h$ on
the convex set $\K_\Xb$ and for any $\Pb \in \K_\Xb$ we get $(\Pb
- \Pb_\Xb) \cdot h^\prime_{\Pb_\Xb} \ge 0$. Besides,
$$
h(\Pb) \ge \Pb \cdot \Vb - l(\Vb)
$$
because $h(\Pb) = \max_{\Vb} { \left\{ \Pb \cdot \Vb - l (\Vb)
\right\} }$. After minimizing we get
$$
h(\Pb_\Xb) \ge \min_{\Pb \in \K_\Xb} { \Pb \cdot \Vb } - l(\Vb) =
\pot^\prime_\Xb (\Vb) - l(\Vb).
$$
But $h(\Pb_\Xb) = \Pb_\Xb \cdot \Vb_\Xb - l(\Vb_\Xb)$ because
$\Vb_\Xb = h^\prime_\Xb$ that gives
$$
h(\Pb_\Xb) = \pot^\prime_\Xb \left( \Vb_\Xb \right) - l(\Vb_\Xb)
$$
together with (\ref{bp}). Comparing the last two displayed
formulas we get
$$
\pot^\prime_\Xb \left( \Vb_\Xb \right) - l(\Vb_\Xb) \ge
\pot^\prime_\Xb (\Vb) - l(\Vb)
$$
that proves the equivalence of our Hamiltonian and Lagrangian
forms of defining the $h$-gradient.

\medskip

\begin{remark}
If the point ${\Pb_\Xb}$ is not defined uniquely the $h$-gradient
does not depend on it. Indeed, if a smooth convex function has the
same value on a segment, its differentials coincide at the points
of the segment.
\end{remark}

\medskip

\begin{theorem}
\label{hdiff} Let the Hamiltonian be the sum of a linear form and
a positive semi-definite quadratic form:
$$
h (\Pb) = h_1 (\Pb) + h_2 (\Pb);
$$
$e$ be a positive semi-definite quadratic form on $\R^m$ and
$\pot: \R^m \to \R$ be a potential with bounded sub-differentials
such that the difference $e - \pot$ is a convex function on
$\R^m$.
%or, more generally,
%\begin{equation}
%\label{ec} \left| \nbl_h{ \left( \pot \right) }\left( \Xb \right)
%\right| = O(\left| \Xb \right|) \mbox{ as } \left| \Xb \right| \to
%+\infty.
%\end{equation}

Then the Cauchy problem
$$
\Tr^+(t) = \nbl_h{ \left( \pot \right) }\left(\Tr(t)\right), \quad
\Tr(0) = \Ab,
$$
where the left side of the differential equation is the one-way
derivative
$$
\Tr^+(t) = \lim_{\lambda \to +0} \frac{\Tr(t + \lambda) - \Tr(t)}
{\lambda},
$$
has a unique global solution which depends continuously on the
initial point $\Ab$ and the potential $\pot$ provided that the
quadratic form $e$ is fixed.

More precisely, it means that there exists a unique trajectory
$\Tr: [0, + \infty) \to \R^m$ satisfying the differential equation
for any $t \ge 0$ and the initial point $\Tr(0) = \Ab$. Moreover,
if the quadratic form $e$ is fixed then the point $\Tr(t)$ depends
continuously on the time $t$, the initial point $\Ab$, and the
potential $\pot$ with respect to the compact-open topology.
\end{theorem}

\medskip

\begin{remark}
Of course, Theorem\,\ref{hdiff} has local variants but it is more
convenient to formulate and prove it globally.
\end{remark}

\medskip

\begin{remark}
Theorem\,\ref{hdiff} looks correct for any smooth convex
Hamiltonian $h$ but the author has failed to find its proof in
this case.
\end{remark}

\section{Proof of Theorem\,\ref{vd}}

\label{tvd}

Theorem\,\ref{vd} follows from Theorem\,\ref{hdiff} applied in the
strip $\R^d \times [0,T]$ of the affine space-time to the
potentials $\potv^\visc$ and $\pot$. Let
$$
\Xb = (\xb,t), \quad \Pb = (\pb,\sigma), \quad \Vb = (\vb,\tau),
$$
$$
h(\pb,\sigma) = |\pb|^2 / 2 + \sigma,
$$
$$
e(\xb,t) = C(T) \, ( |\xb|^2 + t^2) / 2
$$
where $C(T)$ is the constant from Lemma\,\ref{bnd}. According to
Lemma\,\ref{bnd}, the potentials $\potv^\visc$ satisfy the
conditions of Theorem\,\ref{hdiff} in the strip. Besides,
$$
\nbl_h{\potv^\visc} = (\nbl{\potv^\visc},1)
$$
where $\nbl{\potv^\visc}$ is the usual gradient of the smooth
potential $\potv^\visc$. Theorem\,\ref{gp} shows that the limit
potential $\pot$ satisfies the conditions of Theorem\,\ref{hdiff}
as well and the convergence $\potv^\visc \to \pot$ as $\visc \to +
0$ is uniform in the strip. So, the derivative
$\pot^\prime_{\xb,t}$ exists and it remains to show that
\begin{equation}
\label{exh} \nbl_h{\pot} (\xb,t) = (\ub(\xb,t),1)
\end{equation}
where $\ub(\xb,t)$ is the center of the minimal ball containing
the set $k_{\xb,t}$.

\medskip

\begin{remark}
Formally speaking, we cannot apply Theorem\,\ref{hdiff} in the
strip but we can always extend our potentials up to functions on
the space-time satisfying the conditions of Theorem\,\ref{hdiff}.
\end{remark}

\medskip

In order to show the equality (\ref{exh}) we can use the both
forms of the definition $h$-gradient from Section\,\ref{gde}.

\textit{Hamiltonian form:} According to (\ref{fdn}), the set
$\K_{\xb,t}$ of the sub-differentials of the limit potential
$\pot$ at the point $(\xb,t)$ is the convex hull of the set
$$
\left\{ (\pb, \sigma) \; | \; \pb \in K_{\xb,t}, \, |\pb|^2 / 2 +
\sigma + U(\xb,t) = 0 \right\}.
$$
Hence, the Hamiltonian $h (\pb, \sigma) = |\pb|^2/2 + \sigma$
attains its minimum on $\K_{\xb,t}$ at some point $(\ub(\xb,t),
\sigma_\ast)$ where $\ub(\xb,t)$ is the center of the minimal ball
containing the set $K_{\xb,t}$. But substituting into (\ref{fdn})
$\tau = 0$ and comparing with the formula (\ref{fd}) we get that
the sets $k_{\xb,t}$ and $K_{\xb,t}$ has the same convex hull.
Therefore, $\ub(\xb,t)$ is the center of the minimal ball
containing the set $k_{\xb,t}$ as well, and we get
$$
\nbl_h {\pot} (\xb,t) = h^\prime_{\ub(\xb,t), \sigma_\ast} =
(\ub(\xb,t),1).
$$

\textit{Lagrangian form:} The Lagrangian
$$
l(\vb,\tau) = \left\{
\begin{array}{lll}
|\vb|^2/2 & \mbox{if} & \tau = 1   \\
+ \infty        & \mbox{if} & \tau \ne 1
\end{array}
\right.
$$
is the Legendre transformation of the Hamiltonian $h (\pb, \sigma)
= |\pb|^2/2 + \sigma$. This means that the Lagrangian form of the
definition of the $h$-gradient from Section\,\ref{gde}
$$
l(\vb,\tau) - \pot^\prime_{\xb,t} (\vb,\tau) \to \min_{\vb, \tau}
$$
is nothing but the principle (\ref{min}).

\section{Proof of Theorem\,\ref{hdiff}}

According to Theory of ODE, if the potential $\pot$ is smooth then
the Cauchy problem has a unique solution and the boundedness of
the differentials of the potential $\pot$ guarantees that it is
defined globally.

Moreover, our Cauchy problem has a solution if the potential
$\pot$ is homogeneous of the degree $1$ and \textit{concave}. (The
last word means that the function $-\pot$ is convex.) This key
observation is formulated in the following lemma.

\medskip

\begin{lemma}
If $\pot$ is a concave homogeneous function:
$$
\pot (\lambda \Xb) = \lambda \, \pot (\Xb) \quad \forall \;
\lambda \ge 0,
$$
then the trajectory $\Tr: [0, +\infty) \to \R^m$, $\Tr (t) = t \,
\nbl_h {\pot} (0)$ is a solution of the Cauchy problem
$$
\Tr^+(t) = \nbl_h{\pot} \left( \Tr(t) \right), \quad \Tr(0) = 0.
$$
\label{hom}
\end{lemma}

This lemma is true for any smooth convex Hamiltonian; in fact, the
following proof uses its smoothness only, but the convexity is
needed for the uniqueness of the $h$-gradient.

\medskip

\begin{proof}
Firstly, $\K_\Xb \subset \K_0$ for any point $\Xb \in \R^m$
because our potential $\pot$ is concave and homogeneous of the
degree $1$. (Any sub-differential at any point is a
sub-differential at $0$.)

Secondly, if the smooth Hamiltonian $h$ attains its minimal value
on $\K_0$ at a point $\Pb_0 \in \K_0$ then $\Pb_0 \in \K_{\Tr(t)}$
for any $t \ge 0$. It immediately follows from the equality
$$
\pot( \Tr (t)) = \Pb_0 \cdot \Tr(t)
$$
because $\pot(\Xb) \le \Pb_0 \cdot \Xb$. The last equality follows
from (\ref{bp}). Independently on this reference: $\Tr(t) = t
h^\prime_{\Pb_0}$ and for $t \ge 0$
$$
\pot(t h^\prime_{\Pb_0}) = \min_{\Pb \in \K_0} { \Pb \cdot t
h^\prime_{\Pb_0}} = \Pb_0 \cdot t h^\prime_{\Pb_0}.
$$
Indeed, for any $\Pb \in \K_0$ we get $(\Pb - \Pb_0) \cdot
h^\prime_{\Pb_0} \ge 0$ because $\Pb_0$ is a minimum point of the
smooth Hamiltonian $h$ on the convex set $\K_0$.

Therefore, the inclusions $\Pb_0 \in \K_{\Tr(t)} \subset \K_0$
show that $\Pb_0$ is a minimum point of the Hamiltonian $h$ on the
set $\K_{\Tr (t)}$ which implies that $\nbl_h {\pot} \left( \Tr(t)
\right) = \nbl_h {\pot} \left( 0 \right)$.
\end{proof}

\medskip

Let $\Phi_e$ be the space of all potentials $\pot$ with bounded
sub-differentials such that the differences $e - \pot$ are convex
functions (i.\,e., $\Phi_e$ consists of the potentials satisfying
the conditions of the theorem) and
$$
\g: \, \Phi_e \times \R^+ \times \R^m \to \R^m, \quad (\pot, t,
\Ab) \mapsto \g_\pot^t(\Ab)
$$
be the mapping sending a potential $\pot$, a time $t \ge 0$, and
an initial point $\Ab$ to the value of the solution of our Cauchy
problem at the time $t$. In other words, the trajectory
$$
\Tr(t) = \g^t_\pot(\Ab)
$$
is a solution of the Cauchy problem.

A priori, the mapping $\g$ can be many-valued and defined not
everywhere. Let $D(\g) \subset \Phi_e \times \R^+ \times \R^m$ is
the domain of definition of the mapping $\g$.

\medskip

\begin{lemma}
If the Hamiltonian is the sum of a linear form and a positive
semi-definite quadratic form:
$$
h (\Pb) = h_1 (\Pb) + h_2 (\Pb)
$$
then the mapping $\g$ is one-valued and continuous on $D(\g)$ with
respect to the compact-open topology in the space $\Phi_e$.
\label{cont}
\end{lemma}

\medskip

\begin{remark}
In the case $h(\Pb) = |\Pb|^2/2$ the proof of Theorem\,\ref{lmt}
proves, in fact, Lemma\,\ref{cont} as well. The key place is the
inequality (\ref{cvx}).
\end{remark}

\medskip

\begin{proof}
We are showing the continuity of the mapping $\g$ at a point
$(\pot_\ast, t_\ast, \Xb_\ast) \in \Phi_e \times \R^+ \times
\R^m$. Let us consider an open bounded set $K$ consisting the
trajectory
$$
\Tr_\ast(t) = \g_{\pot_\ast}^t(\Xb_\ast), \quad t \in [0, t_\ast]
$$
-- it can always be done because the $h$-gradient of the potential
$\pot_\ast \in \Phi_e$ is bounded. Let an open bounded convex set
$M$ contain the closure $\bar{K}$ of the set $K$ and $\pot$ be any
potential such that
\begin{equation}
|\pot(\Xb) - \pot_\ast(\Xb)| < \varepsilon \quad \forall \, \Xb
\in \bar{M}. \label{tin}
\end{equation}
According to Theorem\,\ref{gp}, the last condition guarantees that
the sub-differentials of all such potentials at all points of
$\bar{K}$ are bounded by a constant $B$.

Let us consider another trajectory
$$
\Tr(t) = \g_{\pot}^t(\Xb), \quad \Xb \in K
$$
and introduce the following notation:
$$
\delta \Tr(t) = \Tr(t) - \Tr_\ast(t), \quad \delta \Trp(t) =
\Trp(t) - \Trp_\ast(t),
$$
where
$$
\Trp(t) = \Pb^\pot_{\Tr(t)}, \quad \Trp_\ast(t) =
\Pb^{\pot_\ast}_{\Tr_\ast(t)}
$$
are the sub-differentials of the potentials $\pot$ and $\pot_\ast$
where the Hamiltonian $h$ attains its minimal values. The key
inequality
$$
\delta \Trp (t) \cdot \delta \Tr(t) \le  2 \varepsilon + 2 \, e
(\delta \Tr(t))
$$
is almost the inequality (\ref{cvx}). Like there we have
$$
\pot_\ast (\Tr) - \pot_\ast (\Tr_\ast) \le \Trp_\ast \cdot (\Tr -
\Tr_\ast) + e(\Tr - \Tr_\ast),
$$
$$
\pot (\Tr_\ast) - \pot (\Tr) \le \Trp \cdot (\Tr_\ast - \Tr) +
e(\Tr - \Tr_\ast).
$$
Adding these inequality and taking into account that
$$
- 2 \varepsilon < \pot_\ast (\Tr) - \pot_\ast (\Tr_\ast) + \pot
(\Tr_\ast) - \pot (\Tr)
$$
according to (\ref{tin}), we get our key inequality.

Our further proof is coordinate. Let $\Xb = (\Xb_1,\Xb_2) \in \R^m
= \R^{m_1} \times \R^{m_2}$, $\Xb_1 \in \R^{m_1}$, $\Xb_2 \in
\R^{m_2}$ be affine coordinates such that in the dual coordinates
$(\Pb_1, \Pb_2) \in \T^\ast$ the Hamiltonian has a canonical form:
$$
h(\Pb_1, \Pb_2) =\Pb \cdot \Vb + |\Pb_2|^2/2, \quad \Vb \in \T.
$$
Then our trajectories satisfy the differential equations
$$
\Tr^+(t) = \Vb + \left( 0, \Trp_2(t) \right), \quad \Tr_\ast^+(t)
= \Vb + \left( 0, {\Trp_\ast}_2(t) \right)
$$
because $h^\prime_P = \Vb \cdot d \Pb + \Pb_2 \cdot d \Pb_2$.
Hence,
$$
\delta \Tr_1^+(t) = 0, \quad \delta \Tr_2^+(t) = \delta \Trp_2(t).
$$
Let $|\Xb_1-{\Xb_\ast}_1| < \alpha_1$, $|\Xb_2-{\Xb_\ast}_2| <
\alpha_2$,
$$
\quad R(t) = |\delta \Tr_2 (t)|^2, \quad 2 \, e(\delta \Tr) \le C
\, |\delta \Tr|^2
$$
for some $C \ge 0$. Then
$$
|\delta \Tr_1(t)| = |\Xb_1-{\Xb_\ast}_1| < \alpha_1
$$
and the above key inequality gives that
$$
\delta \Trp (t) \cdot \delta \Tr(t) \le  2 \varepsilon + C \,
|\delta \Tr(t)|^2.
$$
Hence,
$$
R^+(t) = 2 \, \delta \Trp_2 (t) \cdot \delta \Tr_2(t) \le
$$
$$
\le  4 \, \varepsilon + 2 \, C \, |\delta \Tr(t)|^2 - 2 \, \delta
\Trp_1 (t) \cdot \delta \Tr_1(t) \le
$$
$$
\le  4 \, \varepsilon + 2 \, C \, \alpha_1^2 + 2 \, C \, R(t) + 4
\, B \, \alpha_1.
$$
Solving the differential inequality and taking into account that
$R(0) < \alpha_2^2$, we get:
$$
R(t) < (2 \, \varepsilon + C \, \alpha_1^2 + 2 \, B \, \alpha_1)
\frac{e^{2 C t} - 1}{C} + \alpha_2^2 \, e^{2 C t}.
$$
Therefore, as $(\varepsilon, \alpha_1, \alpha_2) \to +0$
$$
|\delta \Tr_1(t)| \to 0, \quad R(t) \to 0
$$
if $t \le t_\ast$. This proves the continuity when time is fixed:
$t= t_\ast$. But if $t \le t_\ast$ we can use the estimate
$$
|\Tr_\ast(t) - \Tr_\ast(t_\ast)| \le (|\Vb| + B) \, |t-t_\ast|,
$$
and in the case $t \ge t_\ast$ --
$$
|\Tr(t) - \Tr(t_\ast)| \le (|\Vb| + B) \, |t-t_\ast|.
$$
\end{proof}

\medskip

Now, in order to prove Theorem\,\ref{hdiff}, it is enough to show
that $D(\g) = \Phi_e \times \R^+ \times \R^m$. Of course, we can
uniquely define a continuous mapping
$$
\gn: \Phi_e \times \R^+ \times \R^m \to \R^m, \quad \gn |_{D(\g)}
= \g.
$$
Indeed, $D(\g)$ is dense in $\Phi_e \times \R^+ \times \R^m$
because all smooth potentials form a dense subset in $\Phi_e$ --
that is shown with the help of the standard smoothing.

It turns out that any trajectory
$$
\Trn(t) = \gn^t_\pot(\Ab)
$$
is a solution of our Cauchy problem and, therefore, $\gn = \g$. In
order to show that we have to check
$$
\Trn^+(0) = \Ab \; \mbox{ and } \; \Trn^+(t) = \nbl_h{ \left( \pot
\right) } \left( \Trn(t) \right).
$$
Of course, the first equality follows from the obvious fact
$\gn^0_\pot = \id$ which is implied by $\g^0_\pot = \id$. Why is
the differential equation satisfied?

The point is that the class of our Cauchy problems is invariant
with respect to adding constants to potentials, translations of
$\Xb$, positive shifts of $t$, and simultaneous dilations of the
graphs of potentials, $t$, and $\Xb$. Hence, the mappings $\g$ and
$\gn$ are invariant with respect to these transformations as well
because the latter are continuous in the spaces $\Phi_e$, $\R^+$,
and $\R^m$. In the terms of the mappings $\gn^t_\pot: \R^m \to
\R^m$
%(where $\pot \in \Phi_e$ and $t \gn 0$)
it means the following.

\begin{enumerate}
\item Adding constants to potentials: $\gn_{\pot + \cnst}^t =
\gn_\pot^t$.

\item Translations of $\Xb$: $\gn^t_{\pot (\Xb - \Ab)} =
\gn_{\pot(\Xb)}^t + \Ab$.

\item Positive shifts of $t$: $\gn_\pot^{t_1 + t_2} =
\gn_\pot^{t_1} \circ \gn_\pot^{t_2}$.

\item Simultaneous dilations of the graphs of potentials, $t$, and
$\Xb$:
$$
\gn^{\lambda t} _{\lambda \pot (\Xb / \lambda)} (\lambda \Ab) =
\lambda \, \gn_{\pot(\Xb)}^t (\Ab), \quad \lambda \ge 1.
$$
(The last inequality guarantees that $\lambda \pot (\Xb / \lambda)
\in \Phi_e$ if $\pot \in \Phi_e$.)
\end{enumerate}

The invariance property\,$3$ implies that the differential
equation is enough to be checked only for $t=0$. According to the
invariance properties $1$ and $2$, we can suppose that $\pot(0) =
0$ and $\Ab = 0$. Besides, the invariance property\,$4$ implies
that
$$
\gn^{t} _{\lambda \pot (\Xb / \lambda)} (0) = \lambda \, \Trn (t /
\lambda), \quad \lambda \ge 1.
$$
Hence,
$$
\Trn^+(0) = \lim_{\lambda \to + \infty} {\lambda \, \Trn (1 /
\lambda)} = \gn^{1} _{\pot^\prime} (0)
$$
because $\Trn(0) = \Ab = 0$ and
$$
\lambda \pot (\Xb / \lambda) \to \pot^\prime (\Xb) \; \mbox{ in }
\; \Phi_e \; \mbox{ as } \; \lambda \to + \infty
$$
in consequence of $\pot(0) = 0$ and Theorems \ref{conv1} and
\ref{gp}. But
$$
\gn^{1} _{\pot^\prime} = \nbl_h{ \left( \pot \right) } (0)
$$
according to Lemma\,\ref{hom} that completes proving
Theorem\,\ref{hdiff}.

\section{Computations for Figures}

Figure\,\ref{ns} shows how particles move around nodes and end
points of the shock. Figures \ref{b}--\ref{i} demonstrate what can
generically happen with a cluster moving in the shock. In order to
get all of these Figures except the one containing an end point
(on the right of Figure\,\ref{ns}), we use the following procedure
for $k=3$ or $4$.

Let in a neighborhood of some point $(\xb_\ast,t_\ast)$ the limit
potential be presented as the minimum of $k$ smooth solutions of
the Hamilton--Jacobi equation:
$$
\pot(\xb,t) = \min{ \left\{ \pot^1(\xb,t), \dots, \pot^k(\xb,t)
\right\} }
$$
where $\pot(\xb_\ast,t_\ast) = \pot^1(\xb_\ast,t_\ast) = \dots =
\pot^k(\xb_\ast,t_\ast)$ and
$$
\pot_t^i(\xb,t) + |\nbl{\pot^i}(\xb,t)|^2 / 2 + U(\xb,t) = 0
$$
for all $i=1,\dots,k$. Therefore, the derivative $\pot^\prime_\ast
= \pot^\prime_{\xb_\ast, t_\ast}$ of the limit potential at the
point $(\xb_\ast, t_\ast)$ is presented by the formula
$$
\pot^\prime_\ast (\vb,\tau) = \min_{i=1,\dots,k} {\left\{ \pb_i
\cdot \vb - \tau |\pb_i|^2 /2 \right\}} - U_\ast \, \tau
$$
where $\pb_i = \nbl {\pot^i} (\xb_\ast,t_\ast)$ and $U_\ast = U
(\xb_\ast, t_\ast)$. Let us linearize the shock and the motion of
particles around the point $(\xb_\ast, t_\ast)$ considering the
derivative $\pot^\prime_\ast$ instead of the limit potential
itself. The shock of the derivative and the velocities of the
particles are described with the help of the last displayed
formula and Theorem\,\ref{vd} in the following way.

Let any three of the momenta $\pb_1, \dots, \pb_k$ do not belong
to the same line and any four of them do not belong to the same
circle.

\underline{For $\tau < 0$.} If the minimum of the values $|\pb_i
\tau - \vb|$, $i = 1, \dots, k$ is attained only for one index
$i_1$ then the particle $\vb$ is outside of the shock and its
velocity is $\pb_{i_1}$. If this minimum is attained for two
indices $i_1, i_2$ then the particle $\vb$ is being at a smooth
point of the shock and its velocity is the midpoint of the segment
$[\pb_{i_1}, \pb_{i_2}]$. If the minimum is attained for three
indices $i_1, i_2, i_3$ then the particle $\vb$ is being situated
at a node of the shock and its velocity is the center of the
minimal disk containing the momenta $\pb_{i_1}, \pb_{i_2},
\pb_{i_3}$.

\underline{For $\tau > 0$.} If the maximum of the values $|\pb_i
\tau - \vb|$, $i = 1, \dots, k$ is attained only for one index
$i_1$ then the particle $\vb$ is outside of the shock and its
velocity is $\pb_{i_1}$. If this maximum attained for two indices
$i_1, i_2$ then the particle $\vb$ is being at a smooth point of
the shock and its velocity is the midpoint of the segment
$[\pb_{i_1}, \pb_{i_2}]$. If the maximum is attained for three
indices $i_1, i_2, i_3$ then the particle $\vb$ is being situated
at a node of the shock and its velocity is the center of the
minimal disk containing the momenta $\pb_{i_1}, \pb_{i_2},
\pb_{i_3}$. Besides, the center of the minimal disk containing all
of the points $\pb_1 \tau, \dots, \pb_k \tau$ belongs to the shock
and the unique trajectory coming from the origin $(\vb, \tau) =
0$.

\medskip

\begin{remark}
If $\tau > 0$, the shock of the derivative is the Voronoi diagram
of the points $\pb_1 \tau, \dots, \pb_k \tau$. If $\tau < 0$, the
shock is the set being analogous to the Voronoi diagram of the
points $\pb_1 \tau, \dots, \pb_k \tau$ but defined by multiple
maxima of the distance (not minima).
\end{remark}

\medskip

Let $k=3$ and the momenta $\pb_1$, $\pb_2$, and $\pb_3$ do not
belong to the same straight line. Then the shock of the derivative
$\pot^\prime_\ast$ has a node at the center of the circle
containing the points $\tau \pb_1$, $\tau \pb_2$, and $\tau \pb_3$
and the velocity of this node is the center of the circle passing
through the momenta $\pb_1$, $\pb_2$, and $ \pb_3$. But the
velocity of the particle situated at the node at a given time is
the center of the minimal disk containing the momenta $\pb_1$,
$\pb_2$, and $ \pb_3$. Of course, these centers do not always
coincide.

Namely, there are two generic possibilities: the triangle with the
vertices $\pb_1$, $\pb_2$, and $ \pb_3$ can be obtuse or acute. If
the triangle is obtuse then the velocities of the node and the
particle are different and the particle leaves the node.
Otherwise, if the triangle is acute then the velocities coincide
and the particle stays at the node. It is convenient to apply the
above procedure in a frame of reference connected with the node.
In this case the shock does not change with time, $|\pb_1|^2 =
|\pb_2|^2 = |\pb_3|^2$, and we get the shock and velocities shown
in Figure\,\ref{ns} on the left (the triangle is obtuse) and in
the middle (the triangle is acute).

\medskip

\begin{remark}
There is another simple explanation of the difference between
these possibilities. Namely, in a frame of reference connected
with the node the derivative $\pot^\prime_\ast$ has no extremum on
the left and does have a maximum in the middle of the
Figure\,\ref{ns}.
\end{remark}

\medskip

At separate times our triangle can become right -- it is shown in
Figures \ref{b} and \ref{s} on the left.

Let $k=4$, and $\pb_1, \dots, \pb_4$ is a \textit{generic}
configuration of momenta. It means that they do not belong to the
same circle and any three of them do not belong to the same line
and do not form a right triangle. Such generic configurations have
many connected components, the components with four obtuse
triangles are called \textit{totally obtuse}. A component is
called \textit{narrow} if the boundary of the minimal disk
containing all the momenta passes only through two of them. A
component is called \textit{wide} if the boundary of the minimal
disk containing all the momenta passes through three of them. Of
course, any totally obtuse configuration is narrow. Each
component, except the totally obtuse ones, defines one of the
types of behavior of clusters during the fifth or sixth
transitions from Figure\,\ref{dd} -- all these types are shown in
the middle and on the right of Figures \ref{b}--\ref{i}. Namely,
taking a configuration from a connected component we apply the
above procedure which shows the following.

Each triangle formed by three of the momenta defines a node before
or after the transition. If this triangle is acute then there is a
growing cluster at the node. If the configuration is totally
obtuse then there are no clusters at all and we ignore it. If the
convex hull of the momenta is a triangle then the fifth transition
occurs, if the convex hull is a quadrangle -- the sixth one does.
If the configuration is not totally obtuse then there is a cluster
after the transition -- its trajectory comes from the origin. If
the configuration is narrow then this cluster is stable, if it is
wide then the cluster is growing.

It remains to compute the velocities around an end point shown on
the right of Figure\,\ref{ns}. Let
$$
\pot(\xb,t) = \min_{\xi}{ \left\{ F(\xi,\xb,t) \right\} }
$$
where $F$ is a family of smooth solutions of the Hamilton--Jacobi
equation:
$$
F_t(\xi,\xb,t) + |\nbl_\xb {F} (\xi,\xb,t)|^2 / 2 + U(\xb,t) = 0
$$
such that the function $F(\cdot,\xb_\ast,t_\ast)$ has the simplest
minimum from the degenerate ones at the point $\xi = 0$. (In
Singularity Theory this minimum is called $A_3$.) It means:
$$
F(\xi, \xb, t) - F(0, \xb, t) = {}
$$
$$
{} = A a^4 + 2 \sum_{i=1}^k B_i \, a^2 b_i + \sum_{i,j=1}^k C_{ij}
b_i b_j + {}
$$
$$
{} + \alpha(\vb,\tau) \, a + \beta(\vb,\tau) a^2 + \sum_{i=1}^k
\gamma_i(\vb,\tau) \, b_i + \dots,
$$
$$
\xi = (a, b_1, \dots, b_k), \quad \vb = \xb - \xb_\ast, \quad \tau
= t - t_\ast;
$$
where the first part of the right side is a positive definite
quadratic form of $a^2, b_1, \dots, b_k$; $\alpha$, $\beta$,
$\gamma_1$, \dots, $\gamma_k$ are linear forms of $(\vb, \tau)$;
and the dots denote the high order terms. This minimum
representation gives us the following.

\begin{enumerate}
\item In a neighborhood of the point $(\xb_\ast, t_\ast)$ the
shock of the limit potential $\pot$  in the space-time is
approximated by the semi-hyperplane
$$
\alpha(\vb, \tau) = 0,
$$
$$
\det{
\begin{pmatrix}
\beta(\vb, \tau) & \gamma_1(\vb, \tau) & \dots & \gamma_k
(\vb, \tau)\\
B_1 & C_{1 1} & \dots & C_{1 k} \\
\vdots & \vdots & \ddots & \vdots \\
B_k & C_{k 1} & \dots & C_{k k}
\end{pmatrix}
} \le 0.
$$

\item If $\pb_\ast = \nbl_\xb {F} (0,\xb_\ast,t_\ast)$ then
$$
\pot^\prime_{\xb_\ast, t_\ast } (\vb,\tau) = \pb_\ast \cdot \vb -
\tau |\pb_\ast|^2 /2 - U(\xb_\ast,t_\ast) \, \tau.
$$
Hence, Theorem\,\ref{vd} implies that $\ub(\xb_\ast, t_\ast) =
\pb_\ast$ and the space-time vector $(\pb_\ast, 1)$ is tangent to
the trajectory starting at the point $(\xb_\ast, t_\ast)$.

\item Differentiating the Hamilton--Jacobi equation we get:
$$
F_{at} + \nbl_\xb {F} \cdot \nbl_\xb {F_a} = 0, \quad F_{b_i t} +
\nbl_\xb {F} \cdot \nbl_\xb {F_{b_i}} = 0,
$$
$$
F_{aat} + \nbl_\xb {F} \cdot \nbl_\xb {F_{aa}} = - \nbl_\xb {F_a}
\cdot \nbl_\xb {F_a}.
$$
After substituting $\xi = 0$, $\vb=0$, and  $\tau = 0$ these
equalities give:
$$
\alpha(\pb_\ast,1) = 0, \quad \gamma_i(\pb_\ast,1) = 0, \quad
\beta(\pb_\ast,1) < 0.
$$
\end{enumerate}

Therefore, the trajectory starting at the end point $(\xb_\ast,
t_\ast)$ of the shock goes inside it because its tangent vector
$(\pb_\ast, 1)$ belongs to the semi-hyperplane from the first
observation. In order to see that, we have to take into account
the third observation and that
$$
\det{
\begin{pmatrix}
C_{1 1} & \dots & C_{1 k} \\
\vdots & \ddots & \vdots \\
C_{k 1} & \dots & C_{k k}
\end{pmatrix}
} > 0
$$
according to the positive definiteness.

\bibliography{shock}
\bibliographystyle{plain}

\end{document}